\documentclass[apj]{emulateapj}
\usepackage{times}
\usepackage{graphicx}
\usepackage{setspace}
\usepackage{amssymb}
\usepackage{multirow}
\usepackage{color}
\usepackage{wrapfig}
\usepackage{float}
\usepackage{lineno}
\usepackage{amsmath}
\usepackage{url}

\newif\ifAMStwofonts
\AMStwofontstrue

\makeatletter

\newcommand{\Rmnum}[1]{\expandafter\@slowromancap\romannumeral #1@}
\makeatother

\shorttitle{Modeling Pulse Profiles of PSR~B1821$-$24}

\shortauthors{Du et al. 2014}

\slugcomment{Accepted for publication in The Astrophysical Journal}

\begin{document}

\title{Modeling Multi-wavelength Pulse Profiles of Millisecond Pulsar
  PSR~B1821$-$24}

 \author{
Yuanjie~Du\altaffilmark{1},~
%
%
Guojun Qiao\altaffilmark{2},~
Ping~Shuai\altaffilmark{1},~
Xiaomin Bei\altaffilmark{1},
Shaolong Chen\altaffilmark{1},
%
%
Linzhong Fu\altaffilmark{1},
Liangwei Huang\altaffilmark{1},
Qingqing Lin\altaffilmark{1},
Jing Meng\altaffilmark{1},
Yaojun Wu\altaffilmark{1},
Hengbin Zhang\altaffilmark{1}
Qian Zhang\altaffilmark{1},
and~
Xinyuan Zhang\altaffilmark{1}
%
}

 \altaffiltext{1}{Qian Xuesen Laboratory of Space Technology,
 NO. 104 Youyi Road, Haidian district, Beijing
   100094, China; dyj@nao.cas.cn}

 \altaffiltext{2}{School of Physics, Peking University, Beijing
   100871, China}

\begin{abstract}
  PSR B1821$-$24 is a solitary millisecond pulsar (MSP) which radiates
  multi-wavelength pulsed photons. It has complex radio, X-ray and
  $\gamma$-ray pulse profiles with distinct peak phase-separations
  that challenge the traditional caustic emission models. Using the
  single-pole annular gap model with suitable magnetic inclination
  angle ($\alpha=40^\circ$) and viewing angle ($\zeta=75^\circ$), we
  managed to reproduce its pulse profiles of three wavebands. It is
  found that the middle radio peak is originated from the core gap
  region at high altitudes, and the other two radio peaks are
  originated from the annular gap region at relatively low
  altitudes. Two peaks of both X-ray and $\gamma$-ray wavebands are
  fundamentally originated from annular gap region, while the
  $\gamma$-ray emission generated from the core gap region contributes
  somewhat to the first $\gamma$-ray peak. Precisely reproducing the
  multi-wavelength pulse profiles of PSR B1821$-$24 enables us to
  understand emission regions of distinct wavebands and justify pulsar
  emission models.

\end{abstract}


\section{Introduction}

PSR~B1821$-$24 (PSR~B1821$-$24A or PSR~J1824$-$2452A) is a solitary
energetic millisecond pulsar (MSP) first discovered in the globular
cluster M28 \citep{lyne87}. It has a spin period $P=3.05$~ms
and a period derivative $\dot{P} = 1.62\times10^{-18}~{\rm s\,s^{-1}}$
according to the ATNF Pulsar
Catalogue\footnote{http://www.atnf.csiro.au/research/pulsar/psrcat/}
\citep[catalogue version: 1.50;][]{psrcat05}. Due to its largest
intrinsic $\dot{P}$ \citep{phinney93}, PSR~B1821$-$24 is believed to
be the most energetic MSP up to date.

PSR~B1821$-$24 has detectable multi-wavelength emission from radio to
$\gamma$-ray band. It was the first MSP with X-ray pulsed emission
clearly detected by \emph{Advanced Satellite for Cosmology and
  Astrophysics} \citep{saito97}. The X-ray pulsed emission of
PSR~B1821$-$24 has also been observed by other X-ray telescopes.
At the radio band, PSR~B1821$-$24 was the first MSP which had been
observed to go through a microglitch in 2001 March \citep{cog04}. This
could be a useful hint to the conventional glitch theory which should
work for both young and old millisecond pulsars.
Similar to other MSPs (e.g., PSR J0218$+$4232, PSR B1937$+$21 or
B1957$+$20), PSR~B1821$-$24 has radio giant pulses with a steep
spectral index and a high degree of polarization, the phase positions
of these giant pulses are overlapped with its X-ray pulses
\citep{knight06}. In addition, \cite{bilous14} first presented
a 720-2400 MHz phase-resolved radio measurements including polarization
angle curve, spectral indices and giant pulses.  


\citet{1821_fermi} showed that $\gamma$-ray, X-ray and radio pulse
profiles with good signal-to-noise ratio for PSR~B1821$-$24, which
challenge the pulsar caustic models \citep{cheng86,dyks03}. Such
models cannot self-consistently explain the complexity of its pulse
profiles.
%
%
PSR~B1821$-$24 was detected to have strongly linearly polarized radio
emission. With the coherently dedispersed polarization data,
\cite{stairs99} used the rotating vector model (RVM) to fit its polarization
curve, and the inclination angle $\alpha=40.\!\!^\circ 7 \pm
1.\!\!^\circ7$ as well as the impact angle $\beta=40^\circ \pm
10^\circ$ are derived, note that the viewing angle is $\zeta=\alpha+\beta$.

Up to date, the multi-wavelength pulse profiles of PSR~B1821$-$24 have
not been well simulated by current emission models due to its
unprecedent complexity. \cite{venter12} speculated that PSR~B1821$-$24
have plausible phase-aligned radio, X-ray and $\gamma$-ray pulse
profiles which had not been precisely simulated. Furthermore, this
pulsar does not have all peaks occurring at the same phase, according to
the observed results presented by \cite{1821_fermi}.

In this paper, the three-dimension annular gap model is used to study
multi-wavelength ($\gamma$-ray, X-ray and radio) pulse profiles for
PSR~B1821$-$24. In Section 2, we will briefly introduce the definition
of annular gap and core gap; the multi-wavelength pulse profiles for
PSR~B1821$-$24 will be simulated, and the altitudes of
multi-wavelength emission will also be identified. The discussion and
conclusion are shown in Section 3 and Section 4, respectively.

\section{Modeling Multi-wavelength Pulse Profiles for
PSR~B1821$-$24 in the Annular Gap Model}

\subsection{Annular Gap and Core Gap}

\begin{figure*}[!tb]
\begin{center}
\includegraphics[angle=0,width=0.98\textwidth]{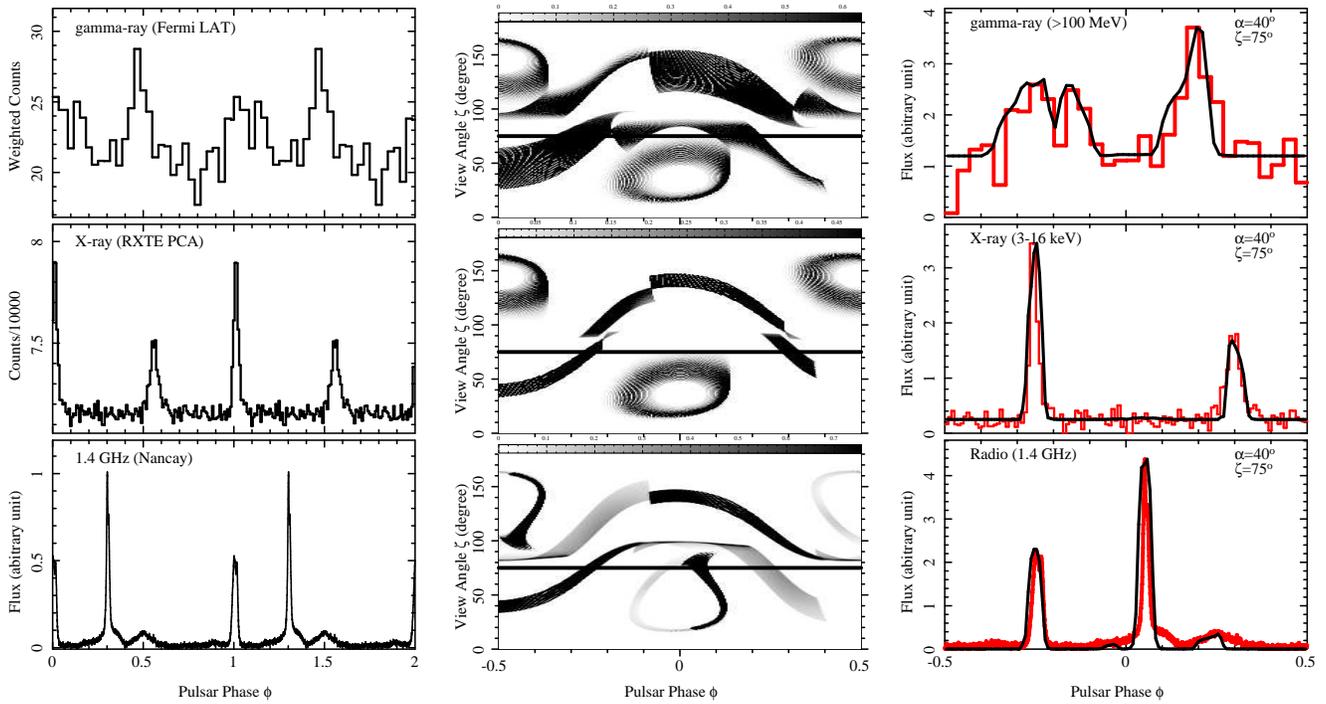}
\caption{Modeled $\gamma$-ray, X-ray and radio pulse profiles for
  PSR~B1821$-$24. Left panels: multi-wavelength observed pulse profile
  data, radio and X-ray data taken from figure 5 of
  \citet{1821_fermi}. Middle panels: two-pole photon sky-map for an
  inclination angle of $\alpha=40^\circ$. Right panels: the
  corresponding simulated pulse profiles (thin black solid lines) cut
  by only one pole emission with a viewing angle of $\zeta=75^\circ$.
  The single-pole annular gap model can well reproduce the observed
  multi-wavelength pulse profiles (thick red solid lines in the right
  panels).
  \\ (A color version of this figure is available in the online
  journal.) \label{fig1} }
\end{center}
\end{figure*}

The polar cap region of a pulsar can be divided into the annular gap
and core gap regions \citep{AG}.
In the vicinity of the light cylinder, charged particles could not
co-rotate with the magnetosphere and they escaped. The pulsar must
output the particles to compensate its loss. Therefore, a huge
acceleration electric field ($E_{\parallel}$) with opposite signs
are generated in both annular gap and core gap regions. This is
the basic physical picture for the generation of $E_{\parallel}$ in the
annular gap model, which is suitable for both millisecond
and young pulsars with short spin periods.
The detailed calculations on $E_{\parallel}$ were presented in
\citep{vela,MSP_AG13}.
A number of primary particles are accelerated to relativistic energies
by $E_{\parallel}$ in the annular and core gaps.
Then the primary particles emanated a great number of $\gamma$-ray photons,
most of them are converted into positron/electron pairs by the processes of
$\gamma$-B and photon-photon annihilation \citep{AG}.
Because of electron/positron pair creation, the plasma in the magnetosphere
could not be fully charge-separated.


\subsection{Multi-wavelength Observations}

When folding $\gamma$-ray or X-ray pulse profiles of a radio-loud
pulsar from \emph{Fermi}-LAT data, a good radio timing solution is
required by the TEMPO2 timing software with some relevant plugins
\citep{hobbs06}. For PSR~B1821$-$24, \cite{1821_fermi} presented a
precise timing solution\footnote{The timing model (par file) of
PSR~B1821$-$24 can be found at
  http://fermi.gsfc.nasa.gov/ssc/data/access/lat/ephems/.}
which is suitable for generating both $\gamma$-ray and X-ray pulse
profiles. Using Tempo2 with the ``Fermi" plugin or ``photons" plugin,
we were able to reproduce the $\gamma$-ray or X-ray pulse profile of
PSR~B1821$-$24. To ensure the validity of the observed pulse profiles,
we directly took the $\gamma$-ray and X-ray profile data from
Figure 5 of \cite{1821_fermi} (see the top and middle figure
in the left panel of Figure~\ref{fig1}). The 1.4 GHz radio profile
data was taken from the webpage of ``Table of Published Ephemerides''
of Fermi Science Support Center.
%
The observed pulse profiles of these three bands are then
renormalized and suppressed to compare with the modeled results by
the annular gap model.

\begin{table*}[!tb]
\centering
\caption{Model parameters of multi-wavelength light curves for PSR~B1821$-$24
  \label{tbl_1}}
%
\begin{tabular}{lcccccccccccccc}
\hline \hline
Band  & $\alpha$ & $\zeta$  & $\kappa$ & $ \lambda$ & $\epsilon$
& $\psi_{\rm cut}$  & $\sigma_{\rm A1}$ &  $\sigma_{\rm A2}$
& $\sigma_{\rm \theta,\,A1}$ & $\sigma_{\rm \theta,\,A2}$ & $\sigma_{\rm C}$
& $\sigma_{\rm \theta,\,C1}$ & $\sigma_{\rm \theta,\,C2}$ \\
\hline
$\gamma$-ray  & 40$^{\circ}$  & 75$^{\circ}$ & 0.94  & 0.8 & 0.8
& 98$^{\circ}$ & 0.2 & 0.1 & 0.002 &  0.001 & 0.24  & 0.0025  & 0.0025 \\
X-ray & 40$^{\circ}$  & 75$^{\circ}$ & 0.84  & 0.8 & 0.8 & 90$^{\circ}$  & 0.2
& 0.2 & 0.001 & 0.001 & 0.24  & 0.0025  & 0.0025 \\
Radio & 40$^{\circ}$  & 75$^{\circ}$ & 0.98  & 0.9 & 0.9  & 120$^{\circ}$ & 0.25
& 0.5   & 0.00375 & 0.0025 & 0.4 & 0.0012  & 0.0025 \\
\hline
\end{tabular}
\end{table*}

The radio profile has three peaks (denoted as peak 1, peak 2 and peak
3 in turn from left to right), while the X-ray and $\gamma$-ray
profiles have two main peaks (denoted as peak 1 and peak 2 from left
to right) with different peak separations. The first radio peak is
almost aligned in phase with that of X-ray and $\gamma$-ray. However,
the other two radio peaks have complex phase separation with
comparison of the second peak of X-ray and $\gamma$-ray
pulses. \cite{1821_fermi} attempted to use the outer-gap model
\citep{cheng86} and the two-pole caustic model \citep{dyks03} to
explain the complex features of multi-wavelength pulse profiles, but
failed, then they drawn a conclusion that it would be difficult to
model the actual pulse profiles across all the three wavebands.
Instead, we will jointly model these three-waveband profiles in our
annular gap model, and reproduce their main features including peaks
and relative phase separation.

\subsection{Modeling Method}

\S3.1 of \cite{vela} presented the detailed method of modeling pulse
profiles.
The parameterized gap width and emissivities were used, which is
similar to the method usually used in the polar cap and slot gap models \citep{harding13}. 
For both of the annular gap and core gap, we project all
radiation from probable emission spots on open filed lines in the
emission region to the resting sky. Meanwhile, both the related
aberration effect and retardation effect are considered.
An assumption of Gaussian
emissivities along an open field line is adopted in the modeling
to speed up calculation, since this numerical assumption is
consistent with physically calculated spectra in the annular gap
model, as presented in Figure 8 of \cite{vela}. Note that emissivities
along different open field lines with a varying magnetic azimuthal
($\psi$) could be different, because the out-flowing particles on
different open field lines (magnetic tubes) with different $\psi$
could have different Lorentz factors and particle numbers. This results
in a patch-like emission pattern (photon sky-map).

The model parameters for modeling radio, X-ray and $\gamma$-ray
pulse profiles of PSR~B1821$-$24 are listed in Table 1. The model
parameters are simply introduced below.
$\alpha$ and $\zeta$ are magnetic inclination angle and viewing
angle; $\kappa$ and $\lambda$ are two geometry parameters used to
scale the peak altitude in the annular gap; $\epsilon$ is a ratio
for the peak altitude in the core gap in units of the peak
altitude in the annular gap; $\psi_{\rm cut}$ is a special
magnetic azimuthal used to constrain different regions which
Gaussian emissivities with different parameters are adopted;
$\sigma_{\rm A1}$ and $\sigma_{\rm A2}$ are length scales in units
of $R_{\rm LC}$ for the emission region on open field lines of
$-180^\circ<\psi_{\rm s}<\psi_{\rm cut}$ and $\psi_{\rm cut}^\circ
\lesssim \psi_{\rm s}<180^\circ$ in the annular gap, respectively;
$\sigma_{\rm C}$ is similar as $\sigma_{\rm A1}$, but a length scale
for the core gap; $\sigma_{\rm \theta,\,A1}$ and $\sigma_{\rm \theta,\,A2}$
as well as $\sigma_{\rm \theta,\,C1}$ and $\sigma_{\rm \theta,\,C2}$
are the transverse bunch scale for field lines of
$-180^\circ<\psi_{\rm s}<\psi_{\rm cut}$ and
$\psi_{\rm cut}^\circ<\psi_{\rm s}<180^\circ$ in the annular gap as
well as core gap, respectively. The more detailed description of
these symbols is presented in \citep{vela}.
We took the values of
$\alpha=40^\circ$ and $\zeta=75^\circ$ which are consistent with those
of the radio polarization measurement of PSR~B1821$-$24 \citep{stairs99}.
$\alpha$ and $\zeta$ are the most important parameters in our model.
When they are chosen, other model parameters which stands for the
emission regions are then carefully adjusted to reproduce the observed
light curve of each band.

\subsection{Modeling Results}

The modeled multi-wavelength pulse profiles of PSR~B1821$-$24 are well
reproduced by the single-pole annular gap model, as presented in
Figure 1. The corresponding emission heights for each band are shown
in Figure 2.
From modeling of the radio pulse, it is found that the radio emission
has a non-caustic origin because of its Gaussian emissivities (see
Table 1), while caustic emission tends to have a constant emissivity
along an field line which is a conventional assumption as used in the
two-pole caustic model \citep{dyks03}. The peak 1 and peak 3 of radio
pulse originate from two regions in the annular gap with intermediate
altitudes of $0.64 R_{\rm LC}$ to $0.85 R_{\rm LC}$ and $0.52 R_{\rm
  LC}$ to $0.78 R_{\rm LC}$, respectively; while radio peak 2 is
generated in the annular gap region with higher altitudes of $0.87
R_{\rm LC}$ to $1.07 R_{\rm LC}$.
For X-ray pulse, it is generated in the annular gap region located at
only one magnetic pole. The emission heights of X-ray peak 1 are
nearly overlapped with those of radio peak 1, while the emission
region of X-ray peak 2 is $0.76 R_{\rm LC}$ to $0.98 R_{\rm LC}$,
slightly higher than the heights of radio peak 3.
For $\gamma$-ray pulse, the emission is also generated in the annular
gap region. Peak 1 is formed by two distinct regions with altitudes
of $0.36 R_{\rm LC}$ to $1.20 R_{\rm LC}$; while peak 2 originates
from a region with altitudes of $0.35 R_{\rm LC}$ to $0.64 R_{\rm
  LC}$.

\begin{figure}[!htb]
\centering
\includegraphics[angle=0,width=0.48\textwidth]{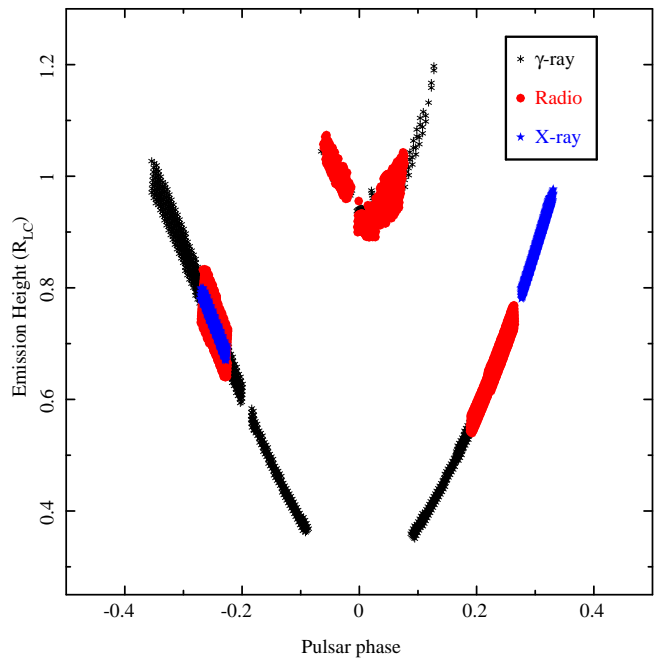}
\caption{Emission heights vesus phase for radio, X-ray and$\gamma$-ray
  pulse profles of PSR~B1821$-$24. It is found that shifted emission
  altitudes for each waveband in the single magnetic-pole, which leads
  to a patch-like emission pattern, can explain the complex properties
  of pulse profiles for PSR~B1821$-$24. }
\label{fig2}
\end{figure}

\section{Discussion}

An observed integrated pulse profile is just a long-time accumulated
photon counting (for X-rays or $\gamma$-rays) or signal power
measurement (for radio emission) based only on a certain orientation
(i.e., line of sight). It is obviously difficult to obtain the whole
emission pattern on the sky for a pulsar. To maximally reproducing the
multi-wavelength profiles of PSR~B1821$-$24, the values of the most
important parameters are chosen as $\alpha=40^\circ$ and
$\zeta=75^\circ$, which are consistent with those derived from
the radio polarization measurement with the RVM.
As noted by \citet{craig14}, the effects of pulsar magnetic field configuration, emission altitude, orthogonal mode jumps, open zone
radius and interstellar scattering would significantly change radio polarization properties, which results in the inaccurate values of
$\alpha$ and $\zeta$ obtained by RVM. Particularly, the situation
will be even more complex when considering the propagation effects
for a radio photon in a pulsar magnetosphere \citep{wang14}. As a
conventional method used by the two-pole caustic model and outer gap
model \citep{dyks03,cheng00}, $\alpha$ and $\zeta$ are are regarded
as free parameters, and obtained via fitting pulse profiles. Instead,
these two values here are adopted from the existing literature \citep{stairs99} to be regarded as a ``reference standard", we carefully
adjusted the values of $\alpha$ and $\zeta$ around the standard until the multi-wavelength pulse profile were reproduced.
Though the final value of $\alpha=40^\circ$ is slightly larger than the conventional values of $\alpha \lesssim 35^\circ$ presented in \cite{AG,MSP_AG13}, but it is still in the allowable range.
%

%
%

Gaussian emissivities with distinct parameters for different region
are also used to simulate profiles for three wavebands, in particular
for the radio band. It is likely that radio emission is originated in
a non-caustic emission mechanism. Radio emission from some pulsars
might be generated in a incoherent style.
For instance, the radio wave could be divided into ordinary mode and
extraordinary mode. When they are propagating in a pulsar's
magnetosphere, due to the refraction effect for the ordinary-mode wave
\citep{wang14}, the outcoming two-mode waves are separated which leads
to a incoherent and non-caustic emissivities.

We predict that other pulsars' multi-wavelength pulse profiles
with complex structures (e.g. PSR J2124$-$3358 with four or more radio
peaks) and peak separations can be well reproduced by the self-consistent
annular gap model with shifted emission altitudes.
Do all normal pulsars (including young and millisecond pulsars) share
the same physical mechanism for radio emission? This is a
long-standing question that needs a larger pulsar sample and further
studies. Joint modeling both radio pulse profile and polarization
angle curve with considerations of various effects discussed above
would be a valuable work which will convincingly constrain the
emission geometry of a pulsar.

\section{Conclusion}

The multi-wavelength (radio, X-ray and $\gamma$-ray) pulse profiles of
PSR~B1821$-$24 can be explained by the annular gap model, and the
emission region of the corresponding band are derived. The pulsed
emission of all the three waveband are generated from the same
magnetic pole. The shifted emission altitudes for each waveband in the
single magnetic-pole, which leads to a patchy emission pattern, can
account for the puzzling properties of pulse profiles for
PSR~B1821$-$24.

\acknowledgments
The authors are grateful to the referee for constructive and helpful comments.
The authors are supported by the National Natural Science Foundation of
China (Grant No. 11303069, 11373011 and 61403391). YJD is also sponsored by
Laboratory Independent Innovation project of Qian Xuesen Laboratory of Space
Technology.


\begin{thebibliography}{}

\bibitem[Bilous et al.(2014)]{bilous14} Bilous, A.~V., Pennucci,
T.~T., Demorest, P., \& Ransom, S.~M.\ 2014, arXiv:1412.7629

\bibitem[Cheng et al.(1986)]{cheng86} Cheng, K.~S., Ho,
  C., \& Ruderman, M.\ 1986, \apj, 300, 500

\bibitem[Cheng et al.(2000)]{cheng00} Cheng, K.~S., Ruderman,
M., \& Zhang, L.\ 2000, \apj, 537, 964

\bibitem[Cognard \& Backer(2004)]{cog04} Cognard, I., \& Backer,
  D.~C.\ 2004, ApJL, 612, L125

\bibitem[Craig(2014)]{craig14} Craig, H.~A.\ 2014, \apj, 790, 102

\bibitem[Du et al.(2010)]{AG} Du, Y.~J., Qiao, G.~J., Han,
  J.~L., Lee, K.~J., Xu, R.~X.\ 2010, \mnras, 406, 2671

\bibitem[Du et al.(2011)]{vela} Du, Y.~J., Han, J.~L., Qiao,
G.~J., \& Chou, C.~K.\ 2011, \apj, 731, 2

\bibitem[Du et al.(2013)]{MSP_AG13} Du, Y.~J., Qiao, G.~J., \& Chen,
  D.\ 2013, \apj, 763, 29

\bibitem[Dyks \& Rudak(2003)]{dyks03} Dyks, J., \& Rudak, B.\ 2003,
  \apj, 598, 1201

\bibitem[Goldreich \& Julian(1969)]{GJ69} Goldreich, P., \& Julian,
  W.~H.\ 1969, \apj, 157, 869

\bibitem[Harding(2013)]{harding13} Harding, A.~K.\ 2013, Journal
of Astronomy and Space Sciences, 30, 145

\bibitem[Hobbs et al.(2006)]{hobbs06} Hobbs, G.~B., Edwards, R.~T., \&
  Manchester, R.~N.\ 2006, \mnras, 369, 655


\bibitem[Johnson et al.(2013)]{1821_fermi} Johnson, T.~J.,
  Guillemot, L., Kerr, M., et al.\ 2013, \apj, 778, 106

\bibitem[Knight et al.(2006)]{knight06} Knight, H.~S., Bailes, M.,
  Manchester, R.~N., \& Ord, S.~M.\ 2006, \apj, 653, 580

\bibitem[Lyne et al.(1987)]{lyne87} Lyne, A.~G., Brinklow, A.,
  Middleditch, J., Kulkarni, S.~R., \& Backer, D.~C.\ 1987, \nat, 328,
  399

\bibitem[Manchester et al.(2005)]{psrcat05} Manchester, R.~N., Hobbs,
  G.~B., Teoh, A., \& Hobbs, M.\ 2005, \aj, 129, 1993

\bibitem[Phinney(1993)]{phinney93}Phinney, E. S. 1993, in ASP
  Conf. Ser. 50, Structure and Dynamics of Globular Clusters,
  ed. S. G. Djorgovski \& G. Meylan (San Francisco, CA: ASP), 141


\bibitem[Saito et al.(1997)]{saito97} Saito, Y., Kawai, N., Kamae, T.,
  et al.\ 1997, ApJL, 477, L37

\bibitem[Stairs et al.(1999)]{stairs99} Stairs, I.~H., Thorsett,
  S.~E., \& Camilo, F.\ 1999, \apjs, 123, 627

\bibitem[Venter et al.(2012)]{venter12} Venter, C., Johnson, T.~J., \&
  Harding, A.~K.\ 2012, \apj, 744, 34

\bibitem[Wang et al.(2014)]{wang14} Wang, P.~F., Wang, C., \& Han,
  J.~L.\ 2014, \mnras, 441, 1943



\end{thebibliography}
\end{document}